\documentclass[prl,aps,twocolumn,showpacs,groupedaddress]{revtex4-1}
\usepackage{color,amsmath,amssymb,amsthm,times,graphics,graphicx,bm,bbm,dcolumn}

\newcommand{\kHz}{{\rm kHz}}

\usepackage{color}

\DeclareMathOperator{\Tr}{Tr}

\begin{document}

\title{Optimal preparation of quantum states on an atom chip device}

\author{
C.\ Lovecchio$^{1}$, F.\ Sch\"{a}fer$^{1,4}$, S.\ Cherukattil$^1$, M.\
Al\`{\i}\ Khan$^1$, I.\ Herrera$^{1,6}$ and F.\ S.\
Cataliotti$^{1,2,3}$,\\ T.\ Calarco$^{5}$, S.\ Montangero$^{5}$, \
F. Caruso$^{1,2,3}$}

\affiliation{$^{1}$LENS and Universit\`{a} di Firenze, Via Nello Carrara 1, 50019 Sesto Fiorentino, Italy}
\affiliation{$^{2}$Dipartimento di Fisica ed Astronomia, Universit\`{a} di Firenze, Via Sansone 1, 50019 Sesto Fiorentino, Italy}
\affiliation{$^{3}$QSTAR, Largo Enrico Fermi 2, 50125 Firenze, Italy}
\affiliation{$^{4}$Department of Physics, Graduate School of Science, Kyoto University, 606-8502 Kyoto, Japan}
\affiliation{$^{5}$Institut f\"{u}r Quanteninformationsverarbeitung \& IQST, Universit\"{a}t Ulm, Albert-Einstein-Allee 11, D-89069 Ulm, Germany}
\affiliation{$^{6}$Centre for Quantum and Optical Science, Swinburne University of Technology, Melbourne, Australia 3122}

\begin{abstract}
Atom chips provide compact and robust platforms towards practical
quantum technologies. A quick and faithful preparation of arbitrary
input states for these systems is crucial but represents a very
challenging experimental task. This is especially difficult when the
dynamical evolution is noisy and unavoidable setup imperfections
have to be considered. Here, we experimentally prepare with very
high small errors different internal states of a Rubidium
Bose-Einstein condensate realized on an atom chip. As a possible
application of our scheme, we apply it to improve the sensitivity of
an atomic interferometer.
\end{abstract}

\maketitle

Microscopic magnetic traps, as atom chips \cite{folman,zimmermann}, represent an important advance in the field of degenerate quantum gases towards their realization of practical quantum technological devices. Indeed, applications outside the laboratory depend on the
compactness and robustness of these setups. Atom chips have already enabled, for instance, the demonstration of miniaturized interferometers \cite{WangPRL05,SchummNP05}, or the creation of squeezed atomic states \cite{RiedelNat10}. Very recently, they have been also applied to the realization of atom interferometers based on non-classical motional states \cite{Frank14}.  Furthermore, atom chips can be integrated with nanostructures \cite{gierling}, or photonic components \cite{kohnen}, to implement new quantum information processing tools, or used as novel platforms for quantum simulations \cite{ivan} and controllable coherent dynamics \cite{QZD}.

As in most quantum technological applications, the ability to quickly and faithfully prepare arbitrary input states is of utmost importance. In particular, it is essential to speedup the initialization protocol of these quantum devices in order to reduce the effects of the inevitably present
noise and decoherence sources. In this context, quantum optimal control provides powerful tools to tackle this problem by finding the optimal way to transform the system from
an experimentally readily prepared initial condition to a desired state with high fidelity from which further quantum manipulations 
can be performed \cite{WR2003,Mabuchi,Rabitz}. It has been successfully implemented
in several physical systems, ranging from cold atoms
\cite{chu2002} to molecules \cite{SB1997}. Recently optimal control algorithms
have been developed and successfully applied to many-body quantum systems
\cite{CRAB,DCM,R2013,jelezko,C2012,C2014}.

In addition, optimal control allows one to speed
up the state preparation process up to the ultimate bound imposed by
quantum mechanics, the so-called quantum speed limit (QSL)
\cite{GLM03,C2009,CCM2011,NP2011}. Indeed, the change of a state into
a different one cannot occur faster than a time scale
that is inversely proportional to the associated energy scale.
This is especially relevant for quantum systems like
Bose-Einstein condensates (BECs) where it is crucial to perform some
desired (coherent) task before dephasing processes
unavoidably occur \cite{Bloch2008}.

\begin{figure}[t]
    \centering
\includegraphics[width=0.49\textwidth,angle=0]{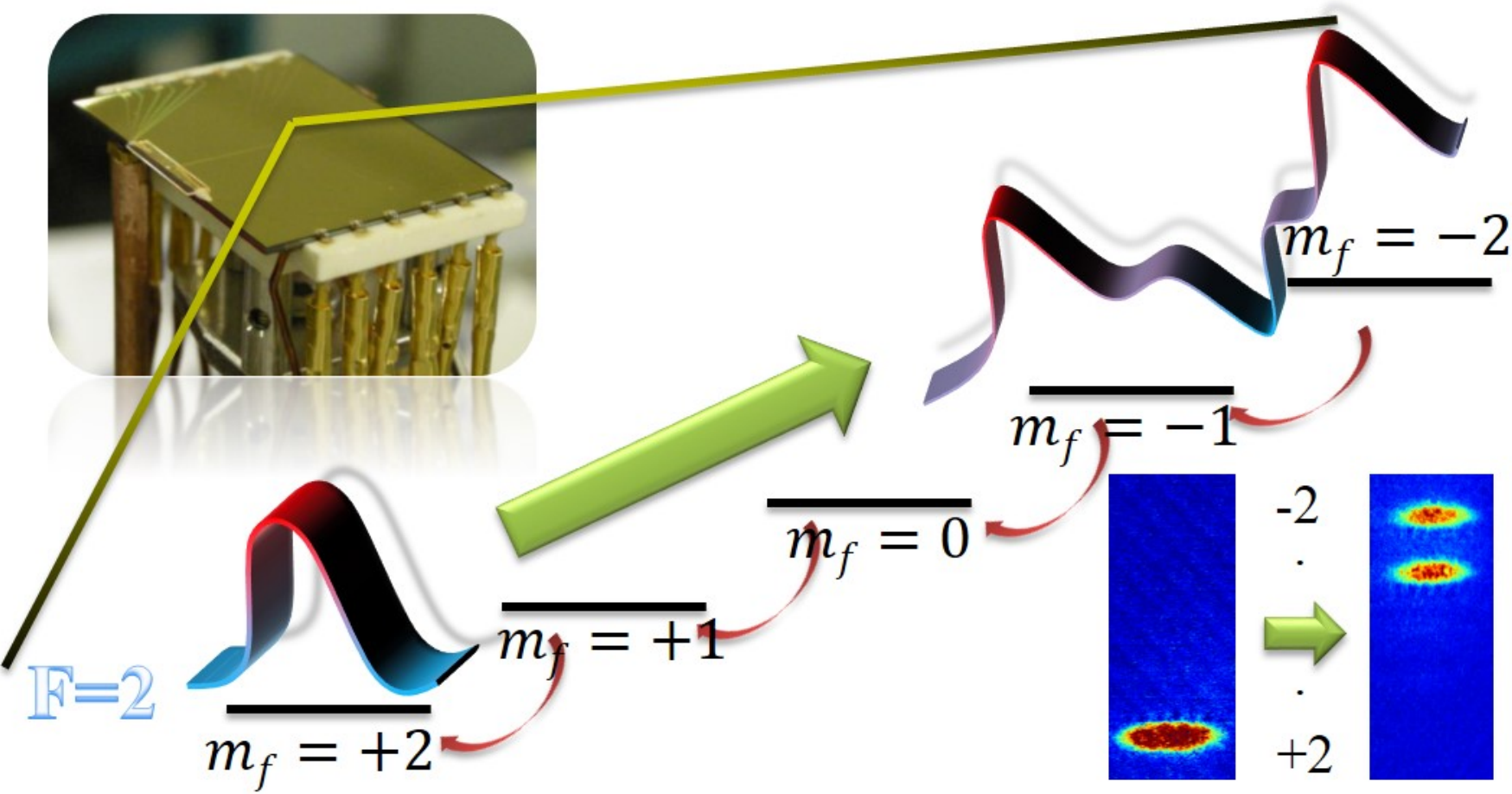}
   \caption{Pictorial representation of the preparation of the internal states of a Bose-Einstein condensate realized on a Atom chip. In the bottom right corner we show the experimentally observed clouds of atoms that are initially in level $+2$ and then halved in the levels $-1$ and $-2$ via optimal control. }
    \label{Fig1}
\end{figure}

In this work we present experimental results obtained on a Rubidium
($^{87}$Rb) BEC, trapped on an atom chip, evolving in a five-level
Hilbert space given by the five spin orientations of the $F=2$
hyperfine ground states --- see Fig.~\ref{Fig1}. We use optimal
control theory to design frequency modulated radio-frequency (RF)
pulses that, starting from a given initial state fixed by the BEC
evaporation procedure, prepare different arbitrary coherent
superpositions of states with high precision. We also show how the
latter scales with the preparation time, with a highly satisfactory
agreement between theory and experiment. We apply, finally, this
technique to improve the sensitivity of an atom interferometer.

\paragraph{Model:}

In the presence of a magnetic field the degeneracy between the spin
orientations of an atomic hyperfine state is lifted. The phenomenon
is described by the Breit-Rabi formula which determines the energies of all different sublevels. For the
five fold $F = 2$ hyperfine ground state of $^{87}$Rb in the presence
of a constant magnetic bias field arbitrarily set to $6.179\rm\, G$ \cite{nota}, the atomic part of the Hamiltonian is $H_0=\hbar \
\text{diag}(\omega_1,\omega_2,\omega_3,\omega_4,\omega_5) =2\pi
\hbar \ \text{diag}(8635,4320,0,-4326,-8657)\rm\, kHz$, where the state basis is chosen to go from $m_F = +2$ to $m_F =
-2$ and the energy values have been
shifted to set $\omega_3=0$.
Under the effect of an RF field it is convenient to write the
evolution of this system in the Rotating Wave Approximation (RWA).
Let us define the terms $V(t)=\hbar \ f(t) \ \text{diag}(-2,-1,0,1,2)$ and
\begin{equation}
H_1=\hbar\left(
    \begin{array}{ccccc}
        0 & \Omega & 0 & 0 & 0 \\
        \Omega & 0 & \sqrt{3/2} \ \Omega & 0 & 0 \\
        0 & \sqrt{3/2} \ \Omega & 0 & \sqrt{3/2} \ \Omega & 0 \\
        0 & 0 & \sqrt{3/2} \ \Omega & 0 & \Omega \\
        0 & 0 & 0 & \Omega & 0
    \end{array}
    \right)\, ,
    \label{eqn:H}
\end{equation}
where $f(t)=\frac{d}{dt} [t \omega(t)]$ describes the energy differences between the five $F=2$ levels within the RWA, $\omega(t)$ is the time dependent frequency of the coupling field that will be modulated in order to implement our optimal control scheme.The Rabi frequency $\Omega$ is proportional to the RF field
intensity and, in all our experiments, we set $\Omega = 2\pi\,60~\kHz$.
Hence, the driving field Hamiltonian is $H_1+V(t)$.

In order to include the unavoidable presence of dephasing noise,
mainly originated in our experiment by the presence of environmental magnetic noise
superimposed on the bias field, we also add in our model
a Lindblad super-operator term ${\cal L}$, acting on the density
matrix $\rho$ as ${\cal L}(\rho) = \sum_{j=1}^{5} \gamma_j [-\{| j
\rangle  \langle j|,\rho\} + 2 | j \rangle  \langle j|\rho | j
\rangle  \langle j|]$, which randomizes the phase of each sublevel $j$ with a rate $\gamma_j$.
For simplicity, we assume the same dephasing rate for
all five sublevels, i.e. $\gamma_j \equiv \gamma$.
The time evolution of the density matrix is then given by
\begin{equation}
\frac{d}{dt} \rho(t) = - \frac{i}{\hbar} [H_0+H_1+V(t),
\rho(t)] + {\cal L}(\rho(t)) \; .
\label{eqn:L}
\end{equation}
\paragraph{Optimal control:}
Here, we use a recently introduced versatile and efficient
optimization algorithm (CRAB)~\cite{CRAB,DCM} for the time dependence
 $f(t)=f_0(1+\sum_{k=-n_f}^{n_f} A_k  \
\exp(i \ \nu_k \ t))$ with $\nu_k=2\pi k/T$, $n_f$ being the number of frequencies,
$A_k$ complex numbers, and $T$ the time duration of the modulation. The optimization is reflected in the modulation of the field's frequency $\omega(t)$,
which we constrain within the range $\omega(t) \in 2\pi [4000, 4700]$ kHz in order
to keep the Rabi frequency $\Omega$ constant (see Supplementary Information for further details). The optimization of the control parameters ($A_k$) in $f(t)$ is done by
means of the Subplex variant of the Nelder-Mead
algorithm~\cite{rowan}. The error function to be minimized is
$\epsilon=\sum_i |\rho_{ii}-b_i|/2$, with $\rho_{ii}$ being the final population
(after time $T$, i.e. at the end of the pulse) of the sublevel $i$,
while $b_i$ are the corresponding target
populations we want to achieve. The initial state $\rho_0$ is given by
all population in the energy level $m_f=+2$, i.e.
$\rho_{11}(t=0)=1$. In the left half of Tab. \ref{table2}, we show the target
populations of different target states (A$,\dots,$I).
Note that there are no constraints on the
coherence terms (i.e., the off-diagonal elements $\rho_{ij}$),
which are however relevant in the context of quantum
state preparation. Even if state tomography would have been in principle
possible \cite{Oberthaler}, it would have been hard to perform it for all
cases studied here. For this reason, in the experiment we measure
only the populations of the five levels, while in the numerical
simulations we take into account the full density matrix
during the optimization. However, we have strong experimental
evidence that the coherence terms (i.e. relative phases) are also
prepared according to our theoretical predictions, because: 1) as
shown in the fine details of the inset of Fig. \ref{Evolution},
there is a very good agreement between theory and experiment in the
population dynamical evolution, 2) when we prepare eigenstates of
the Hamiltonian (see Fig. \ref{HamilChange} and in the Supplementary Information Fig. \ref{SI_Fig2}), they do not evolve anymore, as expected if the Hamiltonian is kept fixed, 3) finally, we consider a
multi-state interferometer and our theoretical results are
reproduced very well by the experimental data. These three tests
would have failed if all the relative phases had not been prepared as in
the target quantum state that is considered in the numerical optimization
protocol.

\paragraph{Experimental setup:}
We have implemented experimentally the optimal control protocols on
a BEC of $^{87}$Rb atoms produced in a
magnetic micro-trap realized on an atom chip. The trap has a
longitudinal frequency of $46~{\rm Hz}$ while the radial trapping
frequency is $950~{\rm Hz}$. The atoms are evaporated to quantum
degeneracy by ramping down the frequency of a radio frequency (RF)
field. The BEC has typically $8\cdot10^4$ atoms, at a critical
temperature of $0.5~\mu{\rm K}$ and is $300~\mu{\rm m}$ from the
chip surface. The magnetic fields for the micro-trap are provided by
a Z-shaped wire on the atom chip and an external pair of Helmholtz
coils. The RF fields for evaporation and manipulation of the Zeeman
states are produced by two additional conductors also integrated on
the atom chip. Finally the homogeneous bias field that lifts the
degeneracy between the spin orientations is created by two further external Helmholtz coils \cite{PetrovicNJP,QZD}.

All manipulations are performed $0.7~\rm{ms}$ after turning off the
magnetic trap to guarantee bias field homogeneity and strongly
reduce the effects of atomic collisions that can produce
decoherence. In this way the most relevant source of noise on the evolution
turns out to be the instability of the environmental magnetic field,
which has a spectrum distribution made of several peaks on a decreasing envelope in the low frequency range. To record the number of atoms in each of the $m_F$ states of the
$F=2$ hyperfine state we apply a Stern-Gerlach method. After the
state manipulation has been performed, in addition to the
homogeneous bias field, an inhomogeneous magnetic field is applied
along the quantization axis for $10~{\rm ms}$. This causes the
different $m_F$ states to spatially separate. After a time of
$23~{\rm ms}$ of expansion a standard absorption imaging sequence is
performed. The atomic populations in each $m_F$ state is normalized to the total number of observed atoms.
\paragraph{Results:}
\begin{table}
\begin{center}
  \begin{tabular}{| c | c | c | c | c | c ||c|c|c|}
    \hline
    \text{Target State} & $\rho_{11}$ & $\rho_{22}$ & $\rho_{33}$ & $\rho_{44}$ & $\rho_{55}$& $\epsilon_{T}$  & $\epsilon_{E}$ & $\cal F$\\\hline
      A   &   1/2 &   0    &   0     & 0 & 1/2&   $0.04(3) $    &   $0.07(1)$     & 0.71\\ \hline
      B   &   1/2 &   0    &   0      & 1/2 & 0 &   $0.04(2)$    &   $0.02(1)$     & 0.67 \\ \hline
      C   &   0 &   1/2  &   0      & 1/2 & 0&   $0.04(3)$   &   $0.04(1)$       & 0.11\\ \hline
      D   &   1/2  &  1/2 &   0   & 0 & 0&   $0.03(2)$   &   $0.02(1)$      & 0.71\\ \hline
      E   &   0  &   1/3  &  1/3 &   1/3 &  0&   $0.04(2)$    &   $0.03(1)$      & 0.02\\ \hline
      F   &   1/5  &   1/5    &   1/5  & 1/5 & 1/5&   $0.02(1)$   &   $0.03(1)$         & 0.45\\ \hline
      G   &   0 &   1    &  0   & 0 & 0  &   $0.05(4)$   &   $0.04(1)$      & 0.15\\ \hline
      H   &   0  &   0   &  0 &  1 &  0&   $0.04(3)$    &   $0.03(1)$      &   0.07\\ \hline
      I   &   0  &  0   &  1 & 0 & 0&   $0.07(3)$    &  $0.07(1)$      & 0.15\\ \hline
  \end{tabular}
       \caption{Target populations $\rho_{ii}$ of different target states (A$,\dots,$I),
       theoretical ($\epsilon_T$) and experimental ($\epsilon_E$) error functions in their preparation,
       and fidelity $\cal F$ between the initial and the target state. The error bars for
       $\epsilon_T$ are obtained considering the maximal deviation when
       the noise rate $\gamma$ is varied in the range $ \gamma \in [20, 200] \ 2\pi$ Hz and the magnetic field
       in a $1$ mG range around the ideal value $B=6.179 \ G$.
       In the absence of dephasing noise, the theoretically predicted $\epsilon_T$ are always smaller than $0.02$.}
       \label{table2}
\end{center}
\end{table}

Let us define the theoretical state preparation error $\epsilon_T$ and the experimentally measured one $\epsilon_E$.  We then apply ten times the same preparation pulse and the reported values of $\epsilon_E$ are the corresponding means and standard deviations.
We take into account the presence of magnetic noise in the experiment by including in the theoretical model an effective dephasing with rate $\gamma \in 2\pi [20, 200]$ Hz (see Eq. \ref{eqn:L}).

\begin{figure}[b]
    \centering
\includegraphics[width=0.5\textwidth,angle=0]{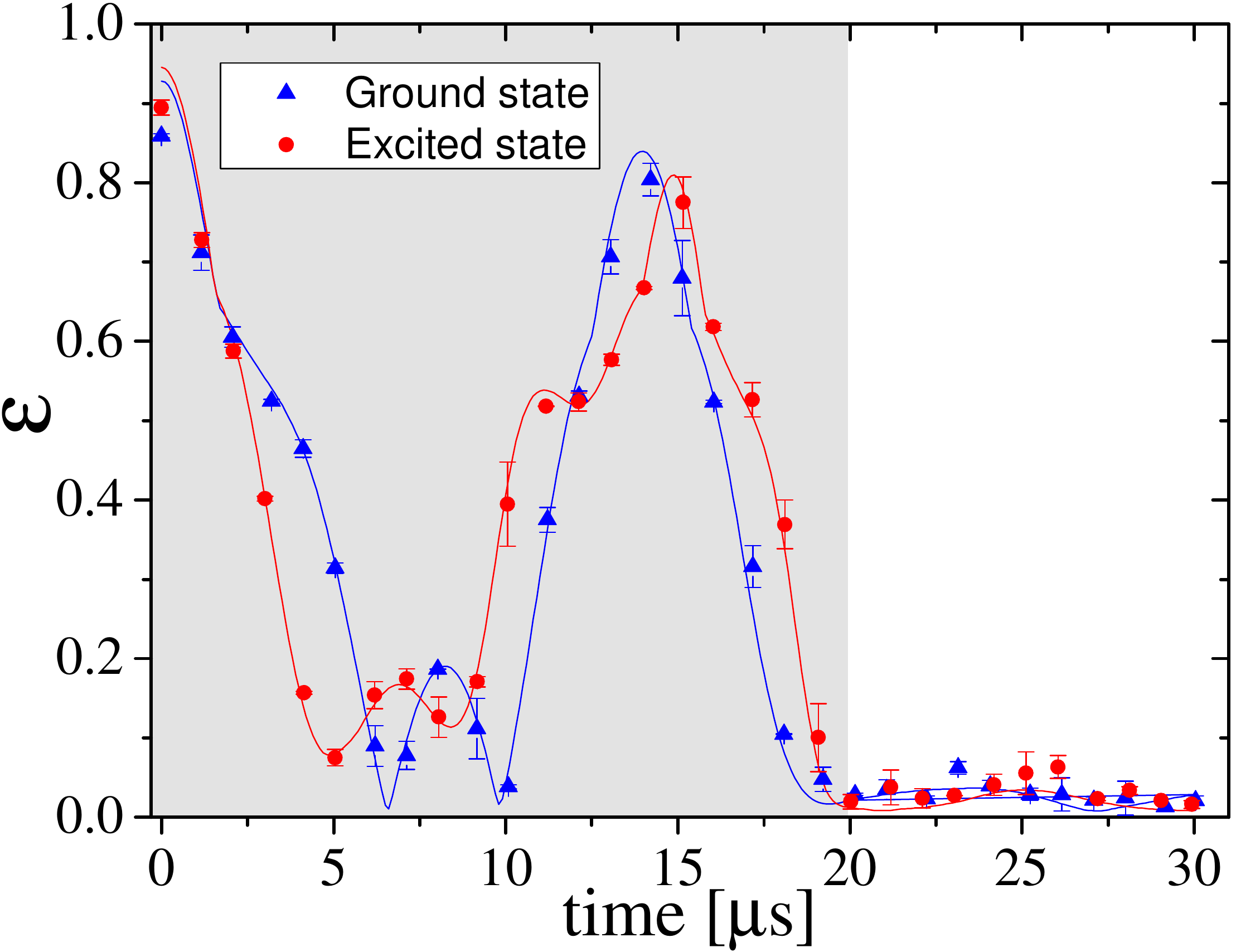}
    \caption{Theoretical (lines) and experimental (dots) time evolution of $\epsilon$ for the preparation (grey area) of a ground state and the highest excited
    state of the Hamiltonian $H_0+H_1+\overline{V}(t)$ with $\overline{V(t)}$ being the harmonic potential that oscillates at $f(t) = \overline{f}=\ 2\pi \  4323$ kHz for every $t \geq T=20 \ \mu s$ (white area).
    }
    \label{HamilChange}
\end{figure}

In the right half of Tab. \ref{table2} we show the errors $\epsilon$ in the preparation of quantum states ($A,\dots,I$),
with a highly satisfactory agreement between theory and experiment. We include
also a measure of distance (Uhlmann fidelity $\cal F$ \cite{nc}) between the initial state $\rho_0$ and the target state $\rho_T$,
which is defined as ${\cal F}(\rho_0,\rho_T)= \Tr \sqrt{\rho_{0}^{1/2} \ \rho_T \ \rho_{0}^{1/2}}$, in order to show that the target states do cover
a large portion of the Hilbert space, i.e. they have varying distances from
the initial state. Let us stress that these states cannot be prepared by simply using constant pulses \cite{PetrovicNJP}.

As noted above, these experimental results only show that we are able to prepare different population distributions in the $F=2$ manifold.
To directly verify our ability to prepare superposition of levels, we have prepared the ground state and highest excited state of the Hamiltonian $H_0 + H_1 + \overline{V}(t)$ in which $f(t)=\overline{f}=\ 2 \pi$ $4323$ kHz, and observed the absence of any evolution of the population distribution of these states after applying the preparation pulse, i.e. for $t \geq T$, while keeping $f(t)=\overline{f}$ ($T=20 \; \mu s$ is the pulse length). 
\begin{figure}[t]
    \centering
\includegraphics[width=0.5\textwidth,angle=0]{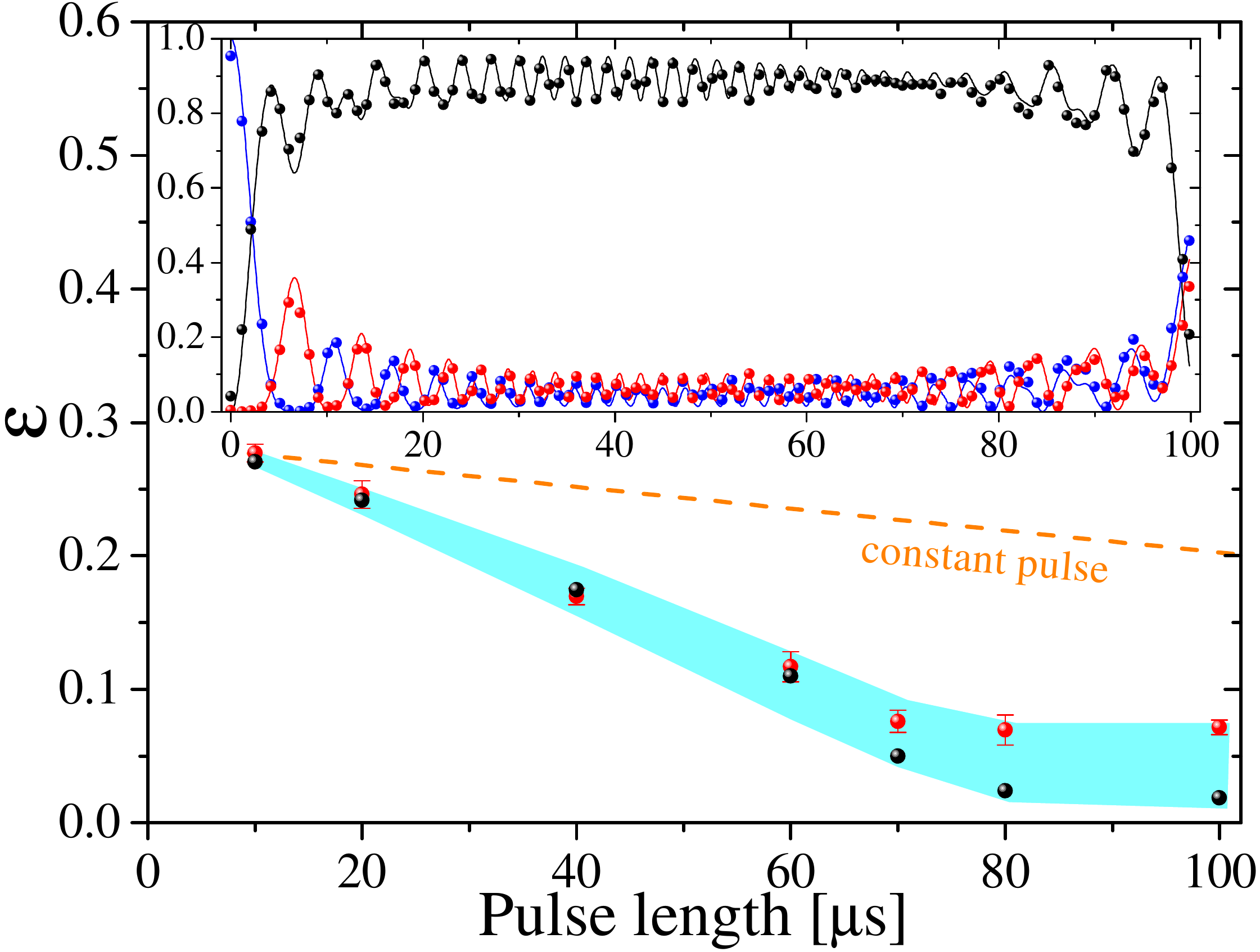}
   \caption{State error preparation $\epsilon$ as a function of pulse length $T$, obtained theoretically (black dots), in the absence of dephasing noise,
   and experimentally (red dots),
   to prepare the state $A$ by optimal control. The effect of
   dephasing noise in the range $2\pi [20, 200]$ Hz is also considered (shaded region).
   The results achievable with constant $f(t) \equiv \overline{f} = 2 \pi \ 4323$ kHz are also shown (dashed line).
   Inset: Theoretical (lines) and experimental (dots) dynamical evolution of the population of the levels $m_F= 2$ (blue), $m_F= -2$ (red), and the sum of the remaining three levels (black) during the application of optimal pulse of length $100 \;\mu s$.
   }
   \label{Evolution}
\end{figure}
The results are reported in Fig. \ref{HamilChange} where we plot the experimental points and expected theoretical behaviour. The theoretical prediction without any free parameter is shown as a continuous line. Some typical examples of optimal pulses are shown in the Supplementary Information.

We then proceed to analyse the time limits for the preparation of the states, in other words
we are interested in keeping the pulse length as short as possible while still suppressing errors
in the state preparation. For this analysis we choose the target state $A$, fixing different pulse times
and minimizing the error function $\epsilon$ under the same constraint on magnetic field and Rabi frequency.
In Fig. \ref{Evolution} we report the results obtained.
The theoretical noiseless results saturate very close to vanishing errors for a pulse length of $90\rm\,\mu s$.
For the experimental results we used the same procedure outlined above to determine the error value.
Experimental results are perfectly matched to the theoretical noiseless value at short pulse times but deviate for longer pulses,
due to the increasing effect of the low frequency experimental noise on the magnetic field.
If we include in the simulation the magnetic field and dephasing noise term fluctuations to correct all the theoretical predictions, we obtain a good agreement with the experimental observation.
The precise matching between theory and experiment is also shown in the inset of Fig. \ref{Evolution} where we report the experimentally measured values of the atomic populations in the five levels during the application of the $100\rm\,\mu s$-pulse. No fitting procedure was used to get the theoretical  results (shown as continuous lines).

\paragraph{Application to interferometry:}
We have implemented an atomic Ramsey interferometer scheme using
some of the state preparation pulses listed above. In a conventional
Ramsey interferometer the application of a $\pi/2$ pulse creates a
balanced superposition of two atomic states separated by an energy
difference $\Delta E$. After the pulse, the atoms are let free to
evolve for a time $\tau$, during which the states acquire a phase
difference $\Delta E/\hbar\, \tau$. Then, the application of a
second $\pi/2$ pulse maps this phase difference onto the atomic
populations, which then oscillate as $\cos(\Delta E/\hbar\, \tau)$.
Even if the smallest signal that an interferometer can ultimately
resolve is limited by the signal to noise ratio, typically its
sensitivity is determined by the maximum slope of the
interferometric fringe. The best achievable sensitivity of a
conventional Ramsey interferometer is found to improve linearly with
$\Delta E$.

The presence of five levels in the $F=2$ manifold complicates the
analysis of the interferometer albeit offering a possibility to
realise a multi-state interferometer with increased sensitivity
\cite{PetrovicNJP}. Whereas optimal control would allow to maximize
the performance of an interferometer \cite{Frank14}, for our system
this would require an extensive analysis of the content of
information of the interferometric output that goes beyond the focus
of the present work. Nonetheless, as an application of our protocol,
we restrict ourselves to interferometric configurations where, for
simplicity, only two levels play a role during the free evolution.
To this aim, we use the state preparation pulses A, B, C, D (all of
them populating two initial states only), both as input and output
beam splitters.

\begin{figure}[t]
    \centering
\includegraphics[width=0.5\textwidth,angle=0]{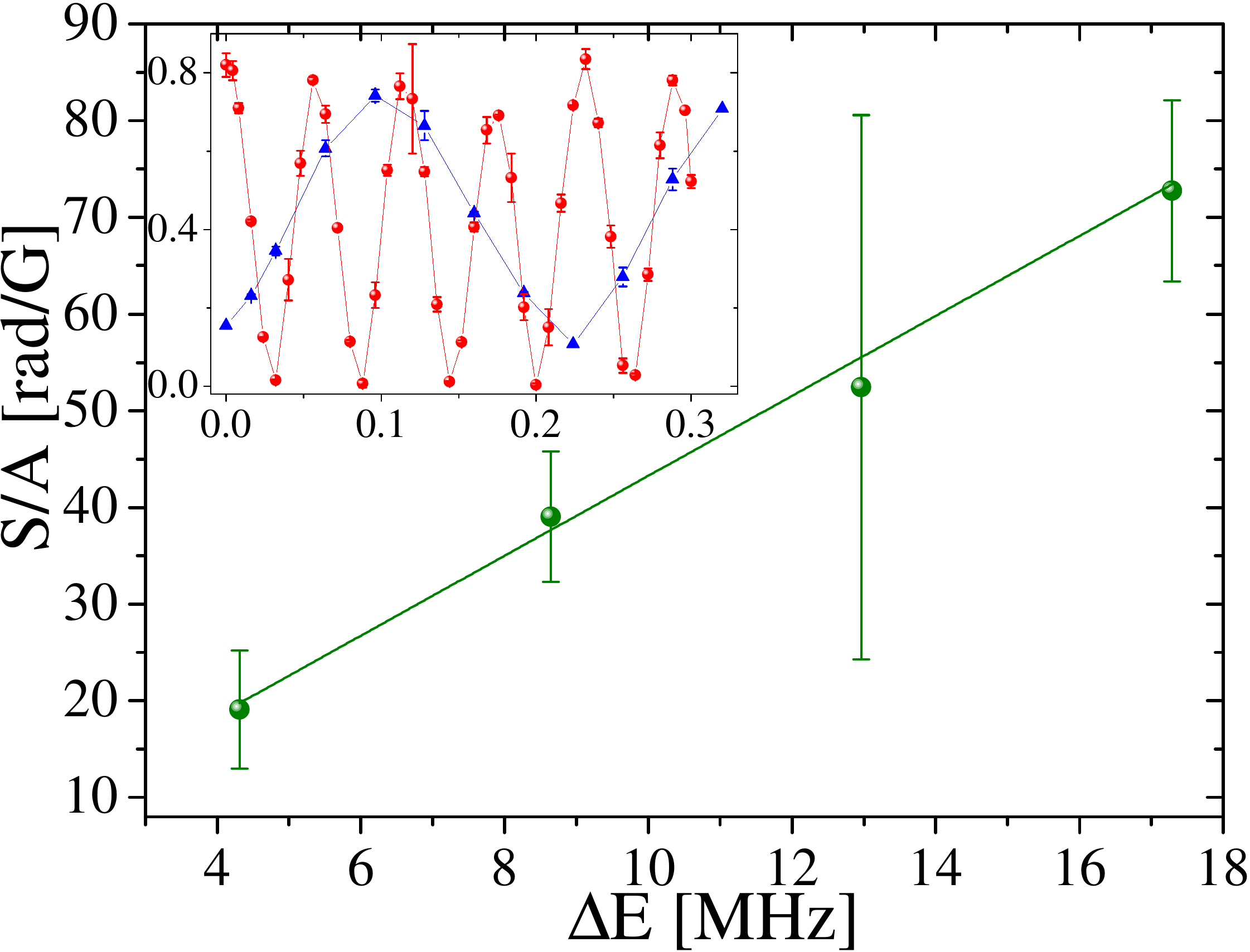}
   \caption{
   Interferometer sensitivity $S$, calculated as the maximum slope of the fringe with the largest amplitude $A$, normalized with respect to the latter. The quantity $S/A$ is plotted as a function of the energy gap $\Delta E$ between the two levels that are initially populated by the optimal control preparation pulse to show
   the linear dependence between them.
   Inset: Relative populations on the level $m_f=+2$ for the state preparation pulses A (red circles) and D (blue triangles) vs. time ($\mu s$).
   }
       \label{Fig4}
\end{figure}

For each preparation pulse we record the interferometric pattern,
i.e. the behaviour of the relative populations in the $F=2$ manifold
after applying the two pulses and a variable time delay $\tau$
between them. Among the five components of the interferometer's
signal we select the one with the highest fringe amplitude $A$ and
we calculate the point-to-point slope selecting its maximum value
$S$. Since the fringe amplitude can be different for each
preparation pulse, we take into account its effect on the
sensitivity by plotting in Fig. \ref{Fig4} the values $S/A$ with
respect to the energy difference $\Delta E$ between the levels
populated via the state preparation pulse. The results confirm,
within the errors, the expected linear dependency between the two
quantities. Let us point out that none of these interferometers can
be realized without optimal control since no one of the states above
(A, B, C, D) can be reached with constant pulses. It must be noted
that, in our experiment, we have not taken into consideration the
fringe amplitude which could be manipulated by a suitable choice of
optimization parameters. However, this simple application still
allows us to achieve a four-fold improvement of the interferometer
sensitivity, without changing neither the applied magnetic field nor
the time interval between the two pulses.

\paragraph{Conclusions:}
The preparation of specific quantum states is a crucial step to
control atom-chip devices, especially in the context of quantum
information and communication technologies. However, this is highly
non-trivial to perform even theoretically, for instance for complex
quantum systems, and especially from the experimental side since one
has to deal with imperfections and the presence of noise due to the
external environment, limiting the observation time scale. Over the
last years, the theory of optimal control has allowed to address
this issue in an efficient way, also when working with fragile
quantum system and non-unitary (noisy) evolutions. Here, we have
applied optimal control tools to preparare a family of different
internal states of a Rubidium Bose-Einstein condensate produced in
an atom chip-based micro-trap. In particular, we theoretically and
experimentally find that these states can be prepared with a very
small error ($\epsilon \sim 0.03$), with very short (control) pulses
successfully overcoming the unavoidable presence of dephasing noise.
Hence, we also test these protocols to implement improved
interferometric configurations that would have been not accessible
without the preparation of some input quantum states by optimal
control. Therefore, these results might pave the way for new schemes
towards, among others, better control of quantum dynamics,
preparation of squeezed states, time-inversion of the system
dynamics, realization of quantum gates, and so more powerful quantum
information protocols.

\
\begin{acknowledgments}
This work was supported by the Seventh Framework Programme for
Research of the European Commission, under FET-Open grant MALICIA, 
QIBEC, SIQS, and by DFS via SFB/TRR21. The work of F.C. has been supported by
EU FP7 Marie-Curie Programme (Career Integration Grant) and by
MIUR-FIRB grant (RBFR10M3SB). We thank M. Inguscio for fruitful discussions and continuous support. QSTAR is the MPQ, LENS, IIT, UniFi Joint Center for Quantum Science and Technology in Arcetri.
\end{acknowledgments}

\begin{figure}[b]
    \centering
\includegraphics[width=0.48\textwidth,angle=0]{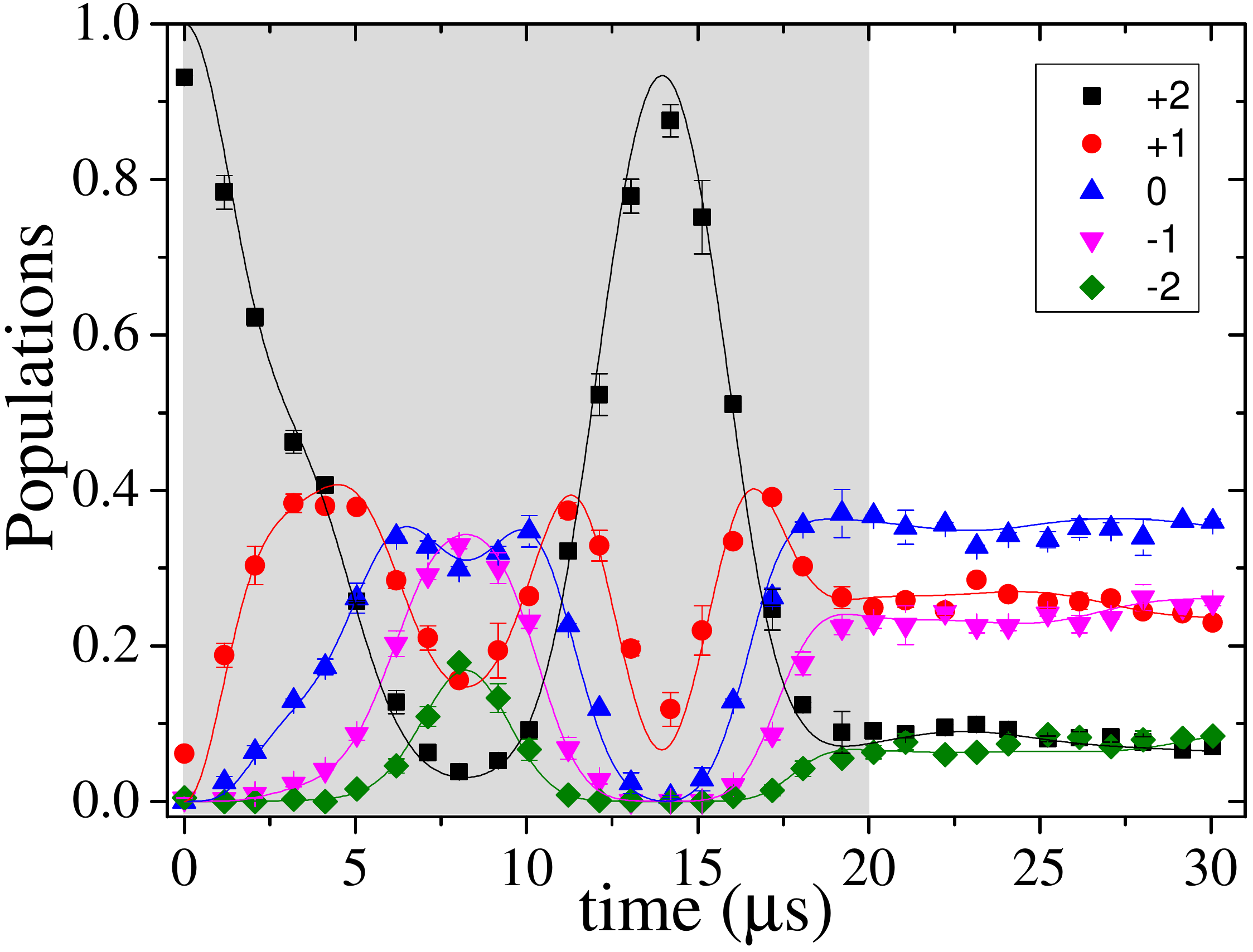}
   \caption{
   Population behaviour of the five sublevels during the time evolution shown in Fig.
   \ref{HamilChange} in the case of the ground state of the Hamiltonian $H_0+H_1+\overline{V}(t)$ with $\overline{V}(t)$ being the harmonic potential that oscillates at $f(t)=\overline{f}=\ 2\pi\; 4323$ kHz, for both theory (lines) and experiment (dots). The gray area correspond to the evolution guided by the optimized pulse, and is followed by a constant frequency driving for every $t \geq T = 20 \  \mu s$, i.e. by the target Hamiltonian.}
    \label{SI_Fig1}
\end{figure}
\section{Supplementary Information}
As introduced in the main text, we use RF induced
evaporation on magnetically trapped $^{87}$Rb atoms to reach the
BEC. The magnetic trap is provided by the chip structures and,
before its loading, we perform optical pumping in the $m_F=2$
magnetic sublevel. This condensation scheme imply that the initial
state of the condensate is the $|F=2,m_F=2\rangle$ state, so as
initial condition every state preparation pulse uses a system in
that pure state. After the BEC creation we let expand the atomic
cloud in the homogeneous magnetic field for $0.7$ ms and then, using
a programmable function generator, we apply to a wire on the chip
the signal $V(t)=V_0 \sin[t \omega(t)]$, which build up the driving
field. In the frequency range $[4000,4700]$ Hz in which we constrain
$\omega(t)$ the radiation efficiency of the used wire can be
considered constant, so the amplitude $V_0$ of the signal can be
determined with high accuracy by performing a series of atomic
population oscillation to have a Rabi frequency $\Omega=2 \pi \;60$
kHz.

To calibrate the homogeneous magnetic field applied by the Helmholtz
coils, we close a multi-state atom interferometer with constant
amplitude pulses as in \cite{PetrovicNJP} from which we deduce the
right current to apply to the coils to produce the value
$B_{th}=6.179 \ G$. To compensate for very slow variations of the
magnetic field we repeat the calibration procedure before the
sequence of ten measurements of a given state preparation pulse.

From a statistical analysis of the interferometer output we also
derive information on the residual noise present in the field. In
particular, by the reduction of the fringe contrast we deduce the
dephasing rate interval $\gamma \in 2 \pi \; \left[20,200\right]$,
while by the variations of the fringe phase we can estimate the
amount of shot-to-shot offset field fluctuations, to $1$ mG.
Introducing these parameters in the theoretical model it is possible
to derive also a maximal theoretical error by taking the worst and
the best output of the model evolution. Those errors are reported in
table \ref{table2}.

To observe the evolution during the optimization pulse, we splice it
in subpulses, measuring the output at the end of each of them.
Figure \ref{SI_Fig1} shows an example of experimental points and
theoretical prediction, reported in solid line, relative to the five
population $m_F=+2 ... -2$ driven from the initial state to the
ground state of the Hamiltonian $H_0+H_1+\overline{V}(t)$ with
$\overline{V}(t)$ being the harmonic potential that oscillates at
$f(t)=\overline{f} = 2\pi \; 4323$ kHz. The optimized preparation
pulse is $20 \mu s$ long and, after the preparation, we keep the
frequency of the potential at the constant target value. The
stability of the populations shows that the system is really in the
target state, with the right coherence between the $|F,m_F\rangle$
components.
\begin{figure}[t]
    \centering
\includegraphics[width=0.48\textwidth,angle=0]{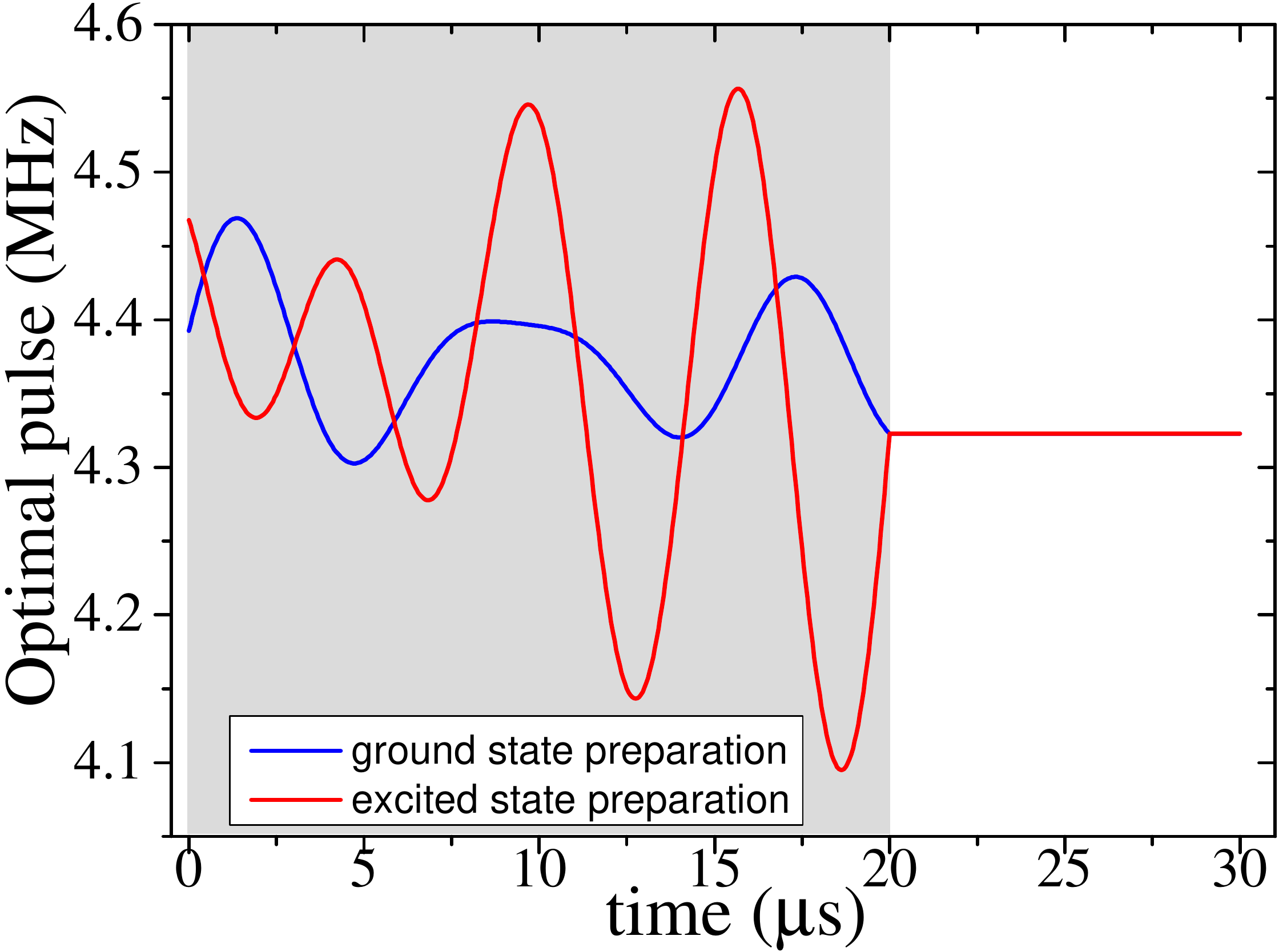}
   \caption{
  Time evolution of the optimally shaped $f(t)/(2\pi)$ for the preparation of the ground state and the excited state of the Hamiltonian $H_0+H_1+\overline{V}(t)$ with $\overline{V}(t)$ being the harmonic potential that oscillates at $f(t)=\overline{f}=\ 2\pi\; 4323$ kHz. The optimally shaped pulse (gray area) is followed by a constant one for $t \geq T = 20 \  \mu s$.}
    \label{SI_Fig2}
\end{figure}
Finally we show in Fig. \ref{SI_Fig2} the optimal pulses for the preparation of the ground and excited states as in Fig.  \ref{SI_Fig1}.
Similar pulses (with $n_f=7$) have been obtained as a result of our optimization scheme to prepare the other states discussed in the main text.


\begin{thebibliography}{99}
\bibitem{nota}
The value of the field $B$ is chosen around $6 \ G$ so that it is much larger than the magnetic noise fluctuations but, at the same time, the current used to produce it is not high enough to cause significant heating of the coils.
\end{thebibliography}
\end{document}